\documentclass[preprint, amsmath,amssymb,aps,natbib,prf]{revtex4-2}
\usepackage{geometry}
\usepackage{comment}
\usepackage{enumerate}
\usepackage{amsmath,nccmath}
\usepackage{xcolor}
\usepackage{mathrsfs}
\usepackage{graphicx}
\usepackage{dcolumn}

\usepackage{natbib}
\usepackage{bm}

\usepackage[mathlines]{lineno}
\usepackage{xcolor}
\usepackage{color}
\usepackage[colorlinks=true,linktoc=page,citecolor=blue,linkcolor=blue,]{hyperref}

\newcommand{\bu}{{\bm u}}

\newcommand{\bomega}{{\bm \omega}}

\newcommand{\bQ}{{\bm Q}}

\newcommand{\bnabla}{{\bm \nabla}}

\begin{document}
\title{Universal energy cascade in homogeneous binary fluid turbulence: A direct comparison of different exact relations}
\author{Nandita Pan}
\email{nandita.pan08@gmail.com}
\author{Supratik Banerjee}
\email{biswayan@gmail.com}
\affiliation{
 Department of Physics, Indian Institute of Technology Kanpur, Uttar Pradesh, 208016, India
}%

\date{\today}
\vspace{-1.5cm}
\begin{abstract}
Below critical temperature, turbulence prevents the spontaneous phase separation of binary mixtures, resulting in a phase arrested state of emulsion which is of significant interest for scientific and industrial applications. The current work is an extensive continuation of our initial theoretical investigation into the nature and universality of associated energy cascade \citep{Pan_2022_Exact_BFT}. In addition to the previously derived divergence and correlator forms of the exact relations, a Banerjee-Galtier type \citep{Banerjee_2016_alternative} divergence-free form of the exact relation is derived under the explicit assumption of homogeneity. By performing three-dimensional direct numerical simulations with up to $1024^3$ grid points, we show that the sum of the kinetic and active energy associates a Kolmogorov-like universal cascade with constant flux rate across inertial scales. Despite term-by-term deviations, the cascade rates computed from all three exact laws show excellent agreement, thereby confirming the equivalence of different exact relations and hence the feasibility of determining the net cascade rate from any of the three formulations. Notably, the two dominant flux rates of the divergence form are found to cross each other roughly at the domain size, which also serves as an infrared cut-off for the inverse cascade of the small scale active energy. This behavior is phenomenologically justified based on the interplay between the flow and surface dynamics of a phase arrested binary fluid.   
\end{abstract}

\maketitle

\begin{centering}
    \section{Introduction}
\end{centering}

 Binary fluids are the mixtures of two immiscible fluids covering the range of simple oil-water mixtures to more complex active fluids \cite{Williamson_acs_2021, Cates_2018_phase-separation_kinetics_active_emulsions, Tiribocchi_2015_ActiveModelH, Pan_2024_Relaxation_BFT, Stalidis_1990_oil-in-water_emulsions, Pine_1984_Turbulent_suppression_Spinodal}.
Above a  critical temperature $T_c$, these two fluids constitute a phase-mixed state, whereas below $T_c$, they tend to phase separate through spinodal decomposition, thus minimizing their interfacial area \cite{Bray_1995_Theory_phase-ordering_kinetics, Cates_2018_phase-separation_kinetics_active_emulsions}. For many practical applications, it is often desirable to suppress phase separation and maintain a large interfacial area even below $T_c$. This is essential in processes such as atomization, froth formation in gas-liquid systems, and emulsification in liquid-liquid systems, where maximizing the interfacial area is a key objective \cite{Hinze_1955}. Emulsions also play a crucial role in various industrial and technological applications, including food production, agrochemicals, pharmaceuticals, materials processing, oil extraction, and mining \cite{Wang_eFood_2023, Shrinivas_2008, Colucci_2020_Water-in-Oil_Emulsions_Delivery_Vehicles, Green_Scilight_crude_oil, Muzzio_1991_self-similar}.
For most binary fluid mixtures, such as simple oil and water, isobutyric acid and water, or 3-methylpentane and nitroethane, $T_c$ is significantly higher than the room temperature. Consequently, these mixtures undergo domain growth or phase-separation under ambient conditions \cite{Williamson_acs_2021,Bailey_2003_3MP_NE,Pine_1984_Turbulent_suppression_Spinodal}. A practical way to prevent such phase-separation is to drive the system through a large-scale turbulent forcing \cite{Berti_2005_Active_Passive_Binary_Mixtures, Perlekar_2014_spinodal_Isotropic_Turbulence, Perlekar_2019_Kinetic_BFT}. Due to its multi-scale mixing properties and aggravated velocity fluctuations, turbulence counteracts the surface tension by fragmenting the large structures of individual fluids and leads to the formation of a homogeneous phase-arrested mixture.\par
In a fully developed turbulent flow, structures of different length scales are nonlinearly coupled with each other and result in an energy cascade from larger to smaller scales until it is dissipated via the viscous effects. Far from both the forcing and dissipation scales, the cascade becomes universal in the sense that it appears to be independent of the large-scale geometry and the small-scale viscous properties. In particular, the universal energy cascade rate $\varepsilon$ becomes practically constant across the inertial length scales $r$. In addition to energy, such cascades are also expected for other inviscid invariants (helicity, enstrophy $etc.,$) of the system. For homogeneous and isotropic hydrodynamic (HD) turbulence, this leads to a universal $k^{-5/3}$ power spectrum for the kinetic energy. In physical space, the quantity $\varepsilon$ can be exactly calculated through the Kolmogorov's $4/5$th exact law as
 $\langle \delta u^3_{r} \rangle = - 4/5 \varepsilon r$, where $\delta u_{r} = \delta \bm{u}\cdot \hat{r}$ is the longitudinal fluctuations of the velocity field $\bu$, and $\delta \bm{u} = \bm{u}(\bm{x}+\bm{r}) - \bm{u}(\bm{x})$ is the two-point velocity fluctuation \cite{von_Karman_1938, kolmogorov1941c}. For homogeneous turbulence which is not necessarily isotropic, one obtains a more general Monin-Yaglom form of the exact laws having a generic form $\bnabla_{\bm{r}}\cdot\bm{\mathcal{F}} = -4 \varepsilon$, where the flux term $\bm{\mathcal{F}} = \langle (\delta \bu)^2 \delta \bu\rangle$ for a HD flow \cite{Monin_Yaglom_2013_Vol2}. As one can verify, Kolmogorov's $4/5$th law can be recovered by simply integrating the generic form under the assumption of isotropy. This divergence form was later extended to other incompressible flows $e.g.,$ passive scalar turbulence, magnetohydrodynamic (MHD) turbulence, and turbulence in stratified flows \cite{Yaglom1949, Politano_1998_dynamical_MHD, Augier_2012_strtified}. Further generalizations to compressible HD and MHD turbulence lead to the exact laws with an additional source term \cite{Galtier_2011_Compressible_sothermal_Turbulence, Banerjee_2013_compressibleMHD, Hellinger_2021, Banerjee_2014_polytropic_turbulence}.
An alternative divergence-free form of the exact relations is recently derived where $\varepsilon$ is expressed in terms of second-order mixed structure functions of fluid variables \cite{Banerjee_2016_alternative, Andres_2019_HD_Numerical, Banerjee_2020_two-fluid_plasma}. Such a form is particularly useful for complex flows $e.g.,$ helical flows, ferrofluid turbulence and compressible turbulence where the divergence form is not straightforward \cite{Banerjee_2016_chiralHMHD, 
Banerjee_2017_self-gravitating_turbulence,
Banerjee_2018_compressible_self-gravitating, Mouraya_2019_ferrohydrodynamic_turbulence, Mouraya_2024_Staionary_nonstationary_cascades, Mouraya_2024_Universal_energy_cascade_critically}. \\
 Several analytical, numerical and experimental studies have been performed to explore various aspects of binary fluid turbulence (BFT) including the onset of turbulence below and above $T_C$, calculation of turbulent diffusivity, evolution of domain growth for simple and active binary fluids, transfer of turbulent kinetic and scalar energy, light-scattering experiments on phase-separating binary fluids, droplet dynamics and distribution in emulsions, turbulent mixing of binary antagonistic fluids, turbulent relaxation of bulk and interfaces, $etc.$ \cite{Ruiz_1981_BFT, aronovitz1984turbulence, Hinze_1955,
 Pine_1984_Turbulent_suppression_Spinodal, Berti_2005_Active_Passive_Binary_Mixtures,Cates_2018_phase-separation_kinetics_active_emulsions, Perlekar_2014_spinodal_Isotropic_Turbulence, Perlekar_2019_Kinetic_BFT, Fan_2016_Csacdes_spinodal_decomposition_BFT, Fan_CHNS_a_case_study, Pan_2024_Relaxation_BFT, Esposito_2023_droplet_dynamics_turbulence_cascade, Mukherjee_Safdari_Shardt_Kenjereš_Van_den_Akker_2019, Bauermann_2025_growing_antagonistic_populations}. The total energy, which the sum of kinetic and active energies, is an inviscid invariant of Cahn-Hilliard-Navier-Stokes (CHNS) equations governing the binary fluid flows and is therefore expected to cascade across the inertial scales \cite{Ruiz_1981_BFT, Bhattacharjee_2021_driven_active_matter}. It is only recently that a divergence form of the exact relation is derived for the total energy cascade in homogeneous BFT \cite{Pan_2022_Exact_BFT}. In the same work, an equivalent form of the exact relation is derived where the energy cascade rate $\varepsilon$ is expressed in terms of two-point correlators. Despite this analytical study, to date, no systematic numerical investigation is carried out to calculate $\varepsilon$ using the derived exact relations and search for a universal cascade of total energy in three dimensional homogeneous BFT. Note that, the inertial range energy cascade rate is equal to the energy dissipation rate for stationary turbulence. Thus, $\varepsilon$ enables the computation of turbulent heating rate directly from the macroscopic field variables without the prior knowledge of microscopic dissipation coefficients. This approach is particularly useful in atmospheric turbulence (often consisting of air-water mixtures) or industrial and laboratory experiments, where extracting macroscopic field variables like the fluid velocity and density are accessible.  \\
 In this article, we conduct pseudo-spectral simulations with $512^3$ and $1024^3$ grid points to look for the universality in homogeneous BFT by calculating the flux-rate from the divergence form of the exact relation. Following \cite{Banerjee_2016_alternative}, we also derive an alternative formulation of the exact law for total energy transfer in BFT. Lastly, the three expressions of $\varepsilon$ obtained from the exact relations, including the one derived from the correlator form in \cite{Pan_2022_Exact_BFT}, are compared. Knowledge of various forms of exact laws has its significance in different scenarios. In particular, the divergence forms are required to bolster the knowledge about the cascade direction (direct or inverse) and power spectra whereas the correlator form is important to obtain a detailed view of different Fourier-space transfer rates associated with $\varepsilon$. The third alternative divergence-free form is particularly useful to study the effect of mean field (if there any) along with gaining insights about the corresponding turbulent relaxed states \cite{Banerjee_PVNLT_2023, Pan_2024_Relaxation_BFT}.   


The paper is organized as follows: the governing equations and the inviscid invariants are discussed in Sec.~\ref{Governing_equations} while Sec.~\ref{alternative_form} consists of the derivation of the alternative form of the exact relation along with a discussion of all the three aforementioned forms of exact relations for energy cascade in homogeneous but not necessarily isotropic BFT. In Sec.~\ref{Numerical details}, we describe the numerical methods and simulation details whereas in Sec.~\ref{Result and discussion}, we present the obtained results and discuss our findings. Finally, we summarize and conclude in Sec.~\ref{Summary and conclusion}.

\begin{centering}
\vspace{0.3cm}
    \section{Governing equations and conservation of energy }\label{Governing_equations}
    \vspace{-0.3cm}
\end{centering} 
Let us take a binary fluid consisting of fluids $A$ and $B$ with densities $\rho_A$ and $\rho_B$, respectively. Below $T_c$, the binary fluid phase separates by minimizing the Landau-Ginzberg type of free energy functional $\mathcal{F} = \int \left[ \frac{a}{2} \phi^2 + \frac{b}{4} \phi^4 + \frac{\kappa}{2} (\bnabla \phi )^2\right]
d\tau$, where $d\tau$ represents the volume element, $\phi = (\rho_A - \rho_B)/(\rho_A + \rho_B)$ is the molecular composition variable, the parameter $a \propto (T-T_c)$ is negative, $b$ is positive and $\kappa = h^2|a|$ with $h$ being the interface-width parameter \cite{chaikin1995,Cates_2018_phase-separation_kinetics_active_emulsions}. As mentioned above, BFT is governed by the CHNS equations where the momentum evolution equation has an additional feedback force $-\phi\bnabla\mu$ due to the phase-separation dynamics, where $\mu = \delta \mathcal{F}/ \delta \phi = a \phi + b \phi^3 - \kappa \nabla^2 \phi$ is the chemical potential. CHNS equations are the simplest diffuse-interface model to describe a binary fluid flow, where the interface between the two fluids is represented by a smoothly varying transition layer \cite{Roccon_phase_field_2023, Soligo_JFE_TurbulentFlows_with_Drops_2021}. The complete set of governing equations for an incompressible binary fluid flow is therefore given by \cite{Cates_2018_phase-separation_kinetics_active_emulsions, Pan_2022_Exact_BFT, Pan_2024_Relaxation_BFT}- 
 \begin{gather}
     \bnabla \cdot \bu = \bm{0}, \label{incompressible}\\
     \partial_{t} \bu =  -(\bm{u}\cdot\bnabla) \bm{u} -   \xi \bQ (\bnabla \cdot \bQ) - \bnabla P + \nu \nabla^2 \bu + \bm{f} \label{Eu},\\
     \partial_{t} \bm{Q} =  -\bnabla(\bu\cdot \bQ) + \mathcal{M} \nabla^2 \bnabla\mu , \label{EQ}
 \end{gather}
where $\bQ = \bnabla \phi$ is the composition gradient field,  $\xi = \kappa/\rho $ is the activity parameter and is equal to $\kappa$ when the total density $\rho = \rho_A+\rho_B$ is normalized to unity, $\bm{f}$ is a large-scale turbulent forcing, $\nu$ is the kinematic viscosity and $\mathcal{M}$ is the mobility coefficient. The feedback term $\xi \bQ (\bnabla \cdot \bQ)$ is coming as a part of the total feedback force $-\phi\bnabla\mu$, while the rest of it is absorbed in the total pressure $P= p + u^2/2 + \phi \mu - \Gamma$, where $p$ is the fluid pressure, $\Gamma$ the homogeneous part of the free energy. The typical numerical values of the relevant parameters are given in Table~\ref{Table1}. As shown in previous studies \cite{Ruiz_1981_BFT, Pan_2022_Exact_BFT}, the total energy $\int E d\tau$ is an inviscid invariant for binary fluid system, where
\begin{equation}
    E = \frac{1}{2} ( u^2 + \xi Q^2) \label{Total_energy},
\end{equation}
where the two terms on the $r.h.s$ represent the kinetic and active energies, respectively. In the presence of energy injection and dissipation, the average energy evolution is given by
\begin{equation}
   \partial_t\langle E \rangle =  \varepsilon_{in} - \varepsilon_{diss},
\end{equation}
where $\langle\cdot\rangle$ represents the ensemble average and is equivalent to the space average assuming ergodicity. The quantities $\varepsilon_{in} = \langle \bu \cdot \bm{f} \rangle$ and $\varepsilon_{diss} = \langle \nu \omega^2 - \mathcal{M} \xi \bQ \cdot \nabla^2 \bnabla \mu\rangle$ represent the average injection and average dissipation rates, respectively, with $\bm{\omega}$ being the vorticity vector. In this case, the average energy can still be conserved when the stationarity is achieved by the mutual balance of $\varepsilon_{in}$ and $\varepsilon_{diss}$. 
 
\vspace{0.5cm}
\begin{centering}
    \section{Different formulations of non-isotropic exact laws}\label{alternative_form}
\end{centering}
 \vspace{-0.2cm}
As discussed in the introduction, the inviscid invariant of the flow is expected to exhibit a universal cascade within the inertial range of length scales. To calculate the energy cascade rate $\varepsilon$ from the exact relation, we first define the two-point correlator for the total energy as
\begin{equation}
    \mathcal{R}_E = \left\langle \frac{\bu\cdot\bu^\prime+\xi\bQ\cdot\bQ^\prime}{2} \right\rangle. \label{Correlator}
\end{equation}
By straightforward algebra, we get \cite{Pan_2022_Exact_BFT} 
\begin{align}
 \partial \mathcal{R}_E = \frac{1}{4}  \bm{\nabla}_{\bm{r}}  \cdot \left \langle \left[  (\delta \bm{u})^2   - \xi (\delta \bm{Q})^2 \right] \delta \bm{u}  \right \rangle + \frac{1}{2} \bm{\nabla}_{\bm{r}}  \cdot \langle \xi (\delta \bm{u}\cdot \delta \bm{Q}) \delta \bm{Q}\rangle + D + F, \label{Partial_RE_div}
 \end{align}
where $F = \langle \bm{f}\cdot \bm{u}' + \bm{f}'\cdot\bm{u} \rangle/2$ is two-point average energy injection rate and $D = D_u + D_Q $ is the two-point average energy dissipation rate with $D_u = \nu\langle\bm{u}^\prime\cdot\nabla^2\bm{u}+ \bm{u}\cdot\nabla^2\bm{u}^\prime\rangle/2 $ and $D_Q =  \mathcal{M}\xi \langle\bm{Q}^\prime\cdot\nabla^2\left(\bm{\nabla}\mu\right) +\bm{Q}\cdot{\nabla^\prime}^2\left(\bm{\nabla}^\prime\mu^\prime\right)\rangle/2 $. Under statistical stationarity, the $l.h.s$ of the above equation vanishes. In the limit of infinite Reynolds number, within the inertial range, the small-scale dissipative effects can be neglected. Furthermore, assuming $\bm{f}$ to be effective only at large scales, the average energy injection rate $F= \langle (\bm{f}'\cdot \bm{u} + \bm{f}\cdot \bm{u}')/2 \approx \langle \bm{f}\cdot \bm{u} + \bm{f}'\cdot \bm{u}' \rangle/2 \approx \langle\bm{f}\cdot \bm{u} \rangle = \varepsilon_{in}\equiv \varepsilon$ and the final exact relation becomes
\begin{equation}
\bm{\nabla}_{\bm{r}}  \cdot \left \langle \left[ (\delta \bm{u})^2   - \xi (\delta \bm{Q})^2 \right] \delta \bm{u}   +2\xi (\delta \bm{u}\cdot \delta \bm{Q}) \delta \bm{Q}\right\rangle = -4\varepsilon.\label{div_form_exact}
\end{equation}
Equivalently, the  $l.h.s.$ of Eq.~\eqref{div_form_exact} can also be expressed in terms of the two-point correlators. The correlator form is beneficial since such exact laws can directly be related to energy budget in Fourier-space \cite{Banerjee_2017_self-gravitating_turbulence}. In the Ref.~\cite{Pan_2022_Exact_BFT}, the correlator form of the exact law is also derived, which takes the form
\begin{equation}
    \mathcal{T}_1 (\bm{r}) + \mathcal{T}_2 (\bm{r}) + \mathcal{T}_3 (\bm{r})=  -\varepsilon,
\end{equation}
where $T_1 (\bm{r}) =- \langle  \bu'\cdot(\bu\cdot\bnabla)\bu + \bu\cdot(\bu'\cdot\bnabla')\bu'\rangle/2$,  $T_{2} = -\xi\langle  (\bu'\cdot\bQ)(\bnabla\cdot\bQ) + (\bu\cdot\bQ')(\bnabla'\cdot\bQ')\rangle/2$ and $T_{3}(\bm{r}) = \xi\langle (\bu'\cdot\bQ')(\bnabla \cdot \bQ) + (\bu\cdot\bQ)(\bnabla' \cdot \bQ') \rangle/2$.
A third form of the exact relation can be derived following the alternative formulation proposed in \cite{Banerjee_2016_alternative}. 
In that case, we obtain
\begin{align}
\partial_t \mathcal{R}_E = & \frac{1}{2}\left\langle \bu'\cdot \partial_t \bu + \bu\cdot \partial_t \bu' + \xi (\bQ' \cdot \partial_t \bQ + \bQ \cdot \partial_t \bQ') \right\rangle  \nonumber\\
= & \frac{1}{2}\left\langle \bu' \cdot \left[(\bu \times \bomega) - \xi \bQ (\bnabla\cdot\bQ)  - \bnabla P_T \right] 
+ \bu\cdot [ (\bu' \times \bomega') - \xi \bQ'(\bnabla'\cdot\bQ^\prime) - \bnabla' P_T'] \nonumber \right.\\
&\left.-  \xi \bQ'\cdot\bnabla(\bu\cdot\bQ)-\xi\bQ \cdot\bnabla'(\bu' \cdot \bQ')\right\rangle+D+F \nonumber \\
=&\frac{1}{2}\left\langle -\delta\bu\cdot \delta(\bu \times \bomega)-\xi [ \bu' \cdot \bQ(\bnabla\cdot\bQ )\nonumber + \bu\cdot\bQ^\prime (\bnabla'\cdot\bQ')\right.\\ &\left.  -(\bu\cdot\bQ)(\bnabla'\cdot \bQ') - (\bu' \cdot \bQ')(\bnabla\cdot \bQ)]\right\rangle+ D+ F \nonumber\\
=& \frac{1}{2}\left\langle -\delta[\bu \times \bomega - \xi  \bQ (\bnabla\cdot\bQ)]\cdot \delta \bu - \xi  \delta (\bu\cdot\bQ)\cdot \delta(\bnabla\cdot\bQ)\right\rangle +D+F, \label{RE02}
\end{align}
where $P_T = P+u^2/2$ is the total pressure. To obtain Eq.~\eqref{RE02}, we use the property of statistical homogeneity 
\begin{equation}
\bm{\nabla} \cdot \langle(\cdot)\rangle = - \bm{\nabla}_{\bm{r}}\cdot \langle(\cdot)\rangle = - \bm{\nabla}^{\prime} \cdot \langle(\cdot)\rangle, \label{Del}
\end{equation}
and the following relations
\begin{align*}
   & (i)\, \langle \bu\cdot (\bnabla^\prime P_T^\prime) \rangle = -\langle P_T^\prime (\bnabla\cdot \bu)\rangle=0,\\
   & (ii) \,\bu\cdot (\bu \times \bomega)=\bu^\prime \cdot (\bu^\prime \times\bomega^\prime)=0,\\
   & (iii) \left\langle\bQ^\prime\cdot\bnabla(\bu\cdot\bQ)\right\rangle = -\left\langle (\bu\cdot\bQ)(\bnabla^\prime\cdot\bQ^\prime) \right\rangle.
\end{align*}
Again under stationarity, in the limit of infinite Reynolds number, neglecting the dissipative terms inside the inertial range, we obtain the exact law,
\begin{equation}
     \left\langle \delta[\bu \times \bomega - \xi  \bQ (\bnabla\cdot\bQ)]\cdot \delta \bu + \xi \delta(\bu\cdot \bQ)\delta (\bnabla\cdot\bQ) \right\rangle = 2\varepsilon. \label{RE3}
\end{equation}
The Eq.~\eqref{RE3} is the alternative divergence-free form for the inertial range energy transfer in homogeneous BFT. Note that, the form of Eq.~\eqref{RE3} is not unique and can also be cast as
\begin{equation}
     \left\langle \delta[\bu \times \bomega - \xi  \bQ (\bnabla\cdot\bQ)]\cdot \delta \bu - \xi \delta[\bnabla(\bu\cdot \bQ)]\cdot\delta \bQ \right\rangle = 2\varepsilon. \label{RE4}
\end{equation}
However, for the analysis, we use the first form which is computationally slightly cost effective \footnote{To calculate $\varepsilon$ from Eq.~\eqref{RE3}, we need to compute $\bu$, $\bomega$, $\bQ$, $(\bnabla\cdot\bQ)$ and $(\bu\cdot\bQ)$, whereas an additional computation of $\bnabla(\bu\cdot\bQ)$ is required for Eq.~\eqref{RE4}.}. 
 Below we list the expressions of all the above exact laws derived assuming homogeneity and stationarity but without any prior assumption of isotropy as
\begin{align}
    \textbf{Divergence form:}\,\, \varepsilon  =&   A_{div}(\bm{r}), \label{div_form}\\
    \textbf{Alternative form:}\,\, \varepsilon =& A_{alt}(\bm{r}),\,\,\text{and} \label{alt_form}\\
    \textbf{Correlator form:}\,\, \varepsilon  
    =& A_{corr}(\bm{r}),\label{corr_form}
\end{align}
where
\begin{align}
    A_{div}(\bm{r}) &= \frac{1}{4}\bnabla_{\bm{r}}\cdot \langle [-(\delta \bu)^2+ \xi (\delta \bQ)^2 ]\delta \bu - 2 \xi(\delta \bu \cdot \delta \bQ) \delta \bQ \rangle,\\
    A_{alt}(\bm{r}) &= \frac{1}{2}\left\langle \delta[\bu \times \bomega - \xi  \bQ (\bnabla\cdot\bQ)]\cdot \delta \bu + \xi \delta(\bu\cdot \bQ)\delta (\bnabla\cdot\bQ) \right\rangle, \label{A_alt_r}\,\,\,\text{and}\\
    A_{corr}(\bm{r}) &=\frac{1}{2}\langle \bu'\cdot(\bu\cdot\bnabla)\bu + \bu\cdot(\bu'\cdot\bnabla')\bu' + \xi(\bu'\cdot\bQ)(\bnabla\cdot\bQ) + \xi(\bu\cdot\bQ')(\bnabla'\cdot\bQ') \nonumber\\
    &- \xi (\bu'\cdot\bQ')(\bnabla \cdot \bQ) - \xi(\bu\cdot\bQ)(\bnabla' \cdot \bQ') \rangle\label{A_corr_r}
\end{align}

\begin{centering}
    \section{Numerical details}\label{Numerical details}
\end{centering}
\vspace{-0.5cm}
We first write the governing equations in terms of the dimensionless starred variables. Using some characteristic length $\ell_0$ (system size), characteristic velocity $u_0$ ($r.m.s.$ velocity) and density $\rho_0 (= \rho_A + \rho_B = 1)$ to normalize the governing Eqs.~\eqref{incompressible}-\eqref{EQ}, we get
 \begin{gather}
     \bnabla^* \cdot \bu^* = \bm{0}, \label{incompressible_starred}\\
     \partial_{t^*} \bu^* =  -(\bm{u}^*\cdot\bnabla^*) \bm{u}^* -   \xi^* \bQ^* (\bnabla^* \cdot \bQ^*) - \bnabla^* P^* + \nu^* \nabla^{*2} \bu^* + \bm{f}^* \label{Eu_starred},\\
     \partial_{t^*} \bm{Q^*} =  -\bnabla^*(\bu^*\cdot \bQ^*) + \mathcal{M^*} \nabla^{*2} \bnabla^*( a^* \phi + b^* \phi^3 - \kappa^* \nabla^{*2}\phi), \label{EQ_starred}
 \end{gather}
 where $\bnabla^* \equiv \ell_0\bnabla$, $\bu^* \equiv \bu/u_0 $, $t^* \equiv (u_0/\ell_0)t$, $\bQ^* \equiv \bnabla^*\phi = \ell_0 \bQ$,  $\xi^* \equiv \xi/(u^2_0\ell^2_0)$, 
$P^* \equiv P/u_0^2$, $\nu* \equiv \nu / (u_0 \ell_0)$, $\bm{f}^* \equiv (\ell_0/u^2_0)\bm{f}$,  $\mathcal{M^*} \equiv (\rho_0 u_0/\ell_0)\mathcal{M}$, $a^* \equiv a / (u^2_0 \rho_0)$, $b^* \equiv b/(u^2_0 \rho_0)$ and $\kappa^* \equiv \kappa/(u^2_0 \ell^2_0 \rho_0)$. For the sake of brevity, we shall omit the starred notation hereinafter. 

\begin{figure}
    \centering
\includegraphics[width=0.85\linewidth]{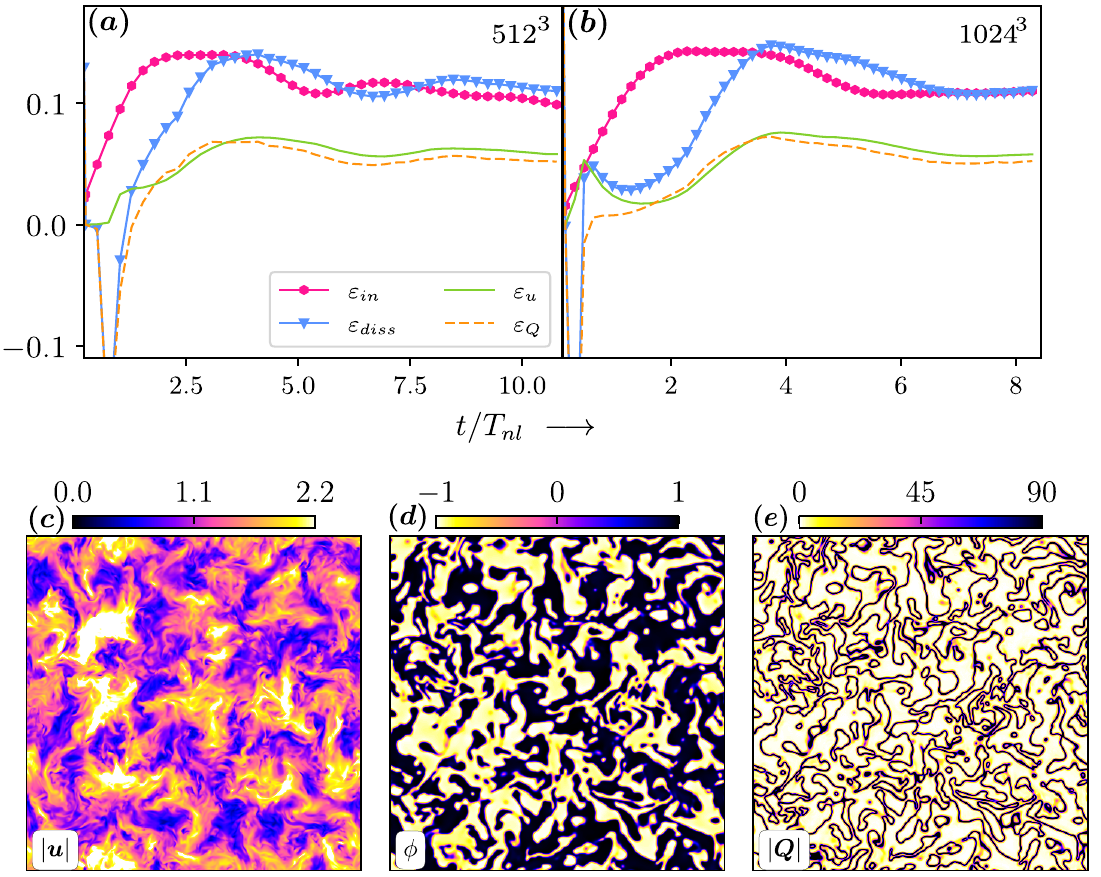}
    \caption{(a), (b) Time evolutions of the average total energy injection ($\varepsilon_{in}$) and dissipation ($\varepsilon_{diss} = \varepsilon_u + \varepsilon_Q$) rates, where $\varepsilon_u$ and $\varepsilon_Q$ correspond to the kinetic energy and active energy dissipation rates, respectively. (c), (d), (e) Snapshots of the modulus of the velocity field ($|\bu|$), the scalar composition field ($\phi$), and the modulus of the composition gradient field ($|\bQ|$), taken at $t = 8T_{nl}$ for Run2.}
    \label{fig:U_phi_Q_Einj_diss}
\end{figure}

The CHNS equations (Eqs.~\eqref{incompressible_starred}-\eqref{EQ_starred}) are simulated using a pseudo-spectral method in a three-dimensional periodic box of volume $(2\pi)^3$ with a total  of $512^3$ and $1024^3$ grid points, respectively. For a simulation with $N^3$ grid points, 
 a $N/2$-dealiasing method is adapted to accommodate the cubic non-linearity present in the chemical potential $\mu$. As a result, the maximum available wavenumber ($k_{max}$) is limited to $N/4$ \cite{canuto2007spectral, Pan_2024_Relaxation_BFT}. The flow is initialized from rest ($\bu(\bm{x},0)=\bm{0}$) with a phase-mixed random distribution of the scalar field $\phi$, where $-0.05<\phi(\bm{x},0)<0.05$. A large-scale Taylor green forcing 
\begin{align}
 \hspace{-0.8cm}  \bm{f} \equiv f_0[sin(k_0x)cos(k_0y)cos(k_0z), - cos(k_0x)sin(k_0y)cos(k_0z),0],
\end{align}
is used to stir the flow where $k_0 =2$ and the forcing magnitude $f_0 = 0.5$.
We used an unconditionally stable explicit-implicit method with an adaptive time-stepping using Courant–Friedrichs –Lewy (CFL) condition \cite{Eyre_1998_Unconditionally_stable_scheme,Yoon_2020_Fourier-Spectral_Method_Phase-Field_Equations}.
A Python-based message-passing interface (MPI) scheme is developed for the parallelization of the code \cite{Pan_2024_Relaxation_BFT, Mortensen_2016_HighPerformance}. Below $T_c$, $\phi$ takes the values $ \phi_{b} = \pm \sqrt{-a/b}$ within the bulk of the individual fluids respectively. Further, choosing $a = -b = -1$, we have $\phi_{b} = \pm 1$ (see Fig.~\ref{fig:U_phi_Q_Einj_diss} c). The kinematic viscosity $\nu$ is chosen in such a way that simultaneously ascertains the high Reynolds number and a well resolved Kolmogorov wavenumber ($k_{\eta}$) $i.e,$ $k_{\eta} =\left(\varepsilon_{u}/\nu^3 \right)^{1/4}<k_{max}$, where $\varepsilon_{u}=\langle \nu \omega^2 \rangle$.
In order to accurately capture the diffuse-interface width $w =4.164 \sqrt{\kappa/|a|}$, the   
parameter $\kappa$ is chosen as $ \kappa = (6\Delta x/4.164)^2|a|$, where $\Delta x = 2\pi/N$ is the minimum grid-spacing \cite{Jacqmin_1999_Two-PhaseNavier–StokesFlowsPhase-FieldModeling, Perlekar_2019_Kinetic_BFT}. In order to achieve high Reynolds number and a wider inertial range, the Kolmogorov-scale $\eta$ is chosen to be small enough. This makes the interface width $w$ and $\eta$ of the same order. However, this does not affect our analysis much as the energy is smoothly dissipated at small-scales and additionally our study is focused on the inertial range transfers. 
The mobility $\mathcal{M} (\propto h)$ is chosen accordingly as the interface thickness reduces, in order to maintain the effective interface force required by the surface tension $\sigma = \sqrt{8 \kappa |a|^3/9b^2}$\cite{ Magaletti_2013_sharp-inerface_limit_of_CHNS, Pan_2024_Relaxation_BFT}. The numerical values of relevant parameters are summarized in Table~\ref{Table1}.

\setlength{\tabcolsep}{7pt} 
\renewcommand{\arraystretch}{1} 
\begin{table*}[ht]
\caption{{\small\label{Table1}{\textbf{Simulation parameters.}} The chosen values of the relevant parameters of the two runs (Run1 and Run2) are listed, where $N^3$ is the total grid points, $\nu$ is the kinematic viscosity, $\mathcal{M}$ the is mobility coefficient, and $\xi$ is the activity parameter. The three important length scales, the integral scale $L \left[= 3\pi/4 \left( \int_0^{\infty} k^{-1} E_u(k) dk/ \int_0^{\infty} E_u(k) dk\right)\right]$, the Taylor length scale $\lambda \left[= \left(5\int_0^{\infty} E_u(k) dk/\int_0^{\infty} k^{2} E_u(k) dk\right)^{1/2}\right]$ and the Kolmogorov scale $\eta \,(=2\pi/k_{\eta})$ are also listed.}}
\vspace{0.5cm}
\begin{tabular}{c c c c c c c c c c c c }
\hline\hline
 Runs  & $N^3$  &   $Re \,(\nu^{-1})$ & $\mathcal{M}$ & $\xi $ & $L$  & $\lambda$ & $\eta$  &  $L/\eta $  &  $k_{max}/k_{\eta}$
\\
\hline
Run1 & $512^3 $ & $1250$ & $0.010$ & $(0.0188)^2$  & $0.537$ & $0.199$ & $0.060$   & $9$ & $1.22$ \\

Run2 & $1024^3 $ & $3300$ & $0.008$ & $(0.0088)^2$  &  $0.526$ & $0.132$ & $0.029$  & $19$ & $1.19$ \\
\hline\hline
\end{tabular}
\end{table*}

\begin{centering}
\section{Results and discussion}\label{Result and discussion}
\end{centering}
\vspace{-0.5cm}
Simulations are evolved over several eddy-turnover times ($ \sim 8T_{nl}$) to ascertain a statistical stationary state where $\varepsilon_{in}$ balances $\varepsilon_{diss}$ (see Fig.~\ref{fig:U_phi_Q_Einj_diss}(a) and (b)).
As expected, the total energy dissipation rate is simply provided by the sum of the dissipation rates of the kinetic and active energies, denoted by $\varepsilon_u$ and $\varepsilon_Q$, respectively.  Note that the kinetic energy dissipation $\varepsilon_u = \langle \nu \omega^2 \rangle$ is a positive definite quantity whereas $\varepsilon_Q = \langle - \mathcal{M} \xi \bQ \cdot \nabla^2\bnabla\mu\rangle$ is not a positive definite quantity. However, for both cases with $512^3$ and $1024^3$ grid points, $\varepsilon_u$ and $\varepsilon_Q$ are found to be fairly equal beyond $ 2 T_{nl}$. To explain this, we remember that both the simulations start from a phase-mixed state where proper-interfaces are not formed. At early stages, the nonlinear term and the higher gradient in $\bQ$ in the chemical potential can be neglected, leading to $\varepsilon_Q = -\mathcal{M}\xi \langle \bQ \cdot \nabla^2 (-\bQ + 3\phi^2\bQ - \kappa \nabla^2\bQ)\rangle \approx -\mathcal{M}\xi \langle -\bQ \cdot \nabla^2 \bQ \rangle \approx -\mathcal{M} \xi \langle (\bnabla\cdot \bQ)^2 \rangle$ a negative value of $\varepsilon_Q$ at initial times (dashed curve in Fig.~\ref{fig:U_phi_Q_Einj_diss} (a) and (b)). 
 At later stages, when the interfaces are formed, the higher gradient term in $\bQ$ shall win over the negative the term and finally $\varepsilon_Q$ becomes positive definite.

In Fig.~\ref{fig:U_phi_Q_Einj_diss} (bottom panel), we plot the snapshots of the dynamical field variables $|\bu|$, $\phi$ and $|\bQ|$. Fig.~\ref{fig:U_phi_Q_Einj_diss}(c) clearly indicates the presence of random flow structures at multiple scales signifying the fully developed turbulent nature of the flow. Additionally, the snapshot also have some localized bright patches representing regions with intense velocity fluctuations. In the absence of turbulent flow, the binary mixture would undergo complete phase separation below $T_c$. However, in the presence of turbulence, a binary fluid attains a phase-locked state of emulsion, where one fluid remains homogeneously suspended with the other, separated by well-defined interfaces (see Fig.~\ref{fig:U_phi_Q_Einj_diss}(d)). In such a state the composition field $\phi$ remains constant inside the bulk of each fluid and only varies across the interface. Therefore, the interface can be delineated by a non-zero value of $|\bQ|$ which vanishes inside the bulk regions (see Fig.~\ref{fig:U_phi_Q_Einj_diss} (e)).

\begin{figure}[h] 
\includegraphics[width=0.9\linewidth]{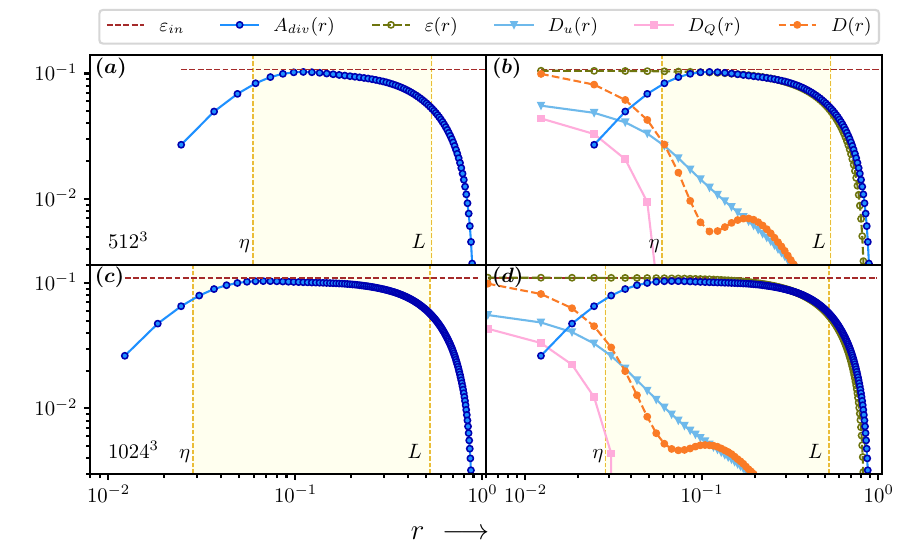}
\caption{(a), (b), (c), (d) Plots of the one-point average energy injection rate $\varepsilon_{in}$, the scale-to-scale nonlinear energy cascade rate $A_{div}(r)$, the average two-point energy injection rate $\varepsilon(r)$, the average two-point total energy dissipation rate $D(r)$, the average two-point kinetic energy dissipation rate $D_u(r)$, and the active energy dissipation rate $D_Q(r)$ as functions of the modulus of the increment vector $r$ for Run1 (top panel) and Run2 (bottom panel). The vertical dashed lines on the left of each subfigure indicate the Kolmogorov scale $\eta$, while those on the right represent the integral scale $L$.}
\label{fig:A_divflux_diss_einj}
\end{figure}

\vspace{-.5cm}
\begin{centering}   \subsection{Universality of energy cascade rate} 
\end{centering}
 \vspace{-0.5cm}
We first numerically investigate if the divergence form can correspond to an inertial range universal energy cascade at a stationary state.  For that, we compute $A_{div}(\bm{r})$ and rewrite it as $A_{div}(\bm{r}) = \bnabla_{\bm{r}}\cdot\bm{\mathcal{F}}(\bm{r}) $ where $\bm{\mathcal{F}}(\bm{r}) = \frac{1}{4}\langle [-(\delta \bu)^2+ \xi (^2\delta \bQ) ]\delta \bu - 2 \xi(\delta \bu \cdot \delta \bQ) \delta \bQ \rangle$. The divergence operator ($\bnabla_{\bm{r}}$) involves the computation of three parts associated with $r$, $\theta$ (polar angle) and $\Phi$ (azimuthal angle) in spherical coordinates where $\bm{r}=(r, \theta, \Phi)$. However, computation of the $\theta$ and $\Phi$ parts of $\bnabla_{\bm{r}}$ is numerically tricky. To overcome this, we adopt the Taylor's angle averaging method where the angular dependence of $\bnabla_{\bm{r}}\cdot\bm{\mathcal{F}}(\bm{r})$ is removed by averaging over $73$ directions spanned by the base vectors $ \in$ \{($1,0,0$), ($1,1,0$), ($1,1,1$), ($2,1,0$), ($2,1,1$) ($2,2,1$), ($3,1,0$), ($3,1,1$)\} on an isotropic sphere \cite{Taylor_2003_Recovering_isotropic_statistics}. Thus, we have 
 $A_{div}(r) = \langle A_{div}(\bm{r})\rangle_{{\theta,\Phi}} = \frac{1} {r^2}\frac{\partial}{\partial r^2}\left(r^2 \langle\mathcal{F}_r \rangle_{\theta,\Phi} (r)\right)$ where $\mathcal{F}_r = \bm{\mathcal{F}}\cdot\widehat{r}$ and $\langle \cdot \rangle_{\theta,\Phi}$ represents the angle averaging.

In Figs.~\ref{fig:A_divflux_diss_einj} (a) and (c), we plot $A_{div}(r)$ as a function of two-point increment $r$ for Run1 ($512^3)$ and Run2 ($1024^3$), respectively. For both the runs, a nearly constant or flat region (shaded region) is found across a range of scales lying between the integral scale ($L)$ and the Kolmogorov scale ($\eta$) representing the sizes of the largest and the smallest possible eddies, respectively. This is in accordance with Kolmogorov's phenomenology, where the intermediate range of scales can be recognized as the inertial range associated with a universal energy cascade having a constant cascade rate. 
For length scales far from $\eta$ but not necessarily far from $L$, $A_{div}(r)$ is found to follow the average two-point injection rate $\varepsilon(r)$ which is not constant across the length scales. However, both $A_{div}(r)$ and $\varepsilon(r)$ become equal to the constant $\varepsilon_{in}=\varepsilon(0)$ for a smaller subset of scales which are also far from $L$ (Figs.~\ref{fig:A_divflux_diss_einj} (b) and (d)) and practically represent the inertial range. This is a direct verification of the exact law derived in Eq.~\eqref{div_form} for a three-dimensional and homogeneous BFT. Further, as the simulation grid point increases from $512^3$ to $1024^3$, the width of the inertial range also increases as observed in the Figs.~\ref{fig:A_divflux_diss_einj} (a) and (c). To explain this, note that the input Reynolds number for Run2 ($Re \sim 3300$) is larger than that of Run1 ($Re \sim 1250$). Further assuming Kolmogorov-type energy cascade due to self-similar fragmentation of eddies, the ratio $L/\eta$ roughly becomes $9$ and $19$ for Run1 and Run2, respectively thus entailing in a wider inertial range in the latter case.  For the sake of clarity, we explicitly plot the two-point average dissipation rates for the kinetic energy, active energy and total energy denoted by $D_u(r)$, $D_Q(r)$ and $D(r)$, respectively (Figs.~\ref{fig:A_divflux_diss_einj} (b) and (c)). Unlike $D_Q(r)$, which becomes negligibly small at and above $\eta$, $D_u(r)$ remains non-negligible around $\eta$ but steadily decreases beyond that. 
The total dissipation rate $D(r)$ therefore  becomes practically small (by at least one-order) w.r.t. $\varepsilon(r)$ thereby assuring an inertial range energy cascade in BFT.

\begin{figure}[ht!]
\includegraphics[width=0.85\linewidth]{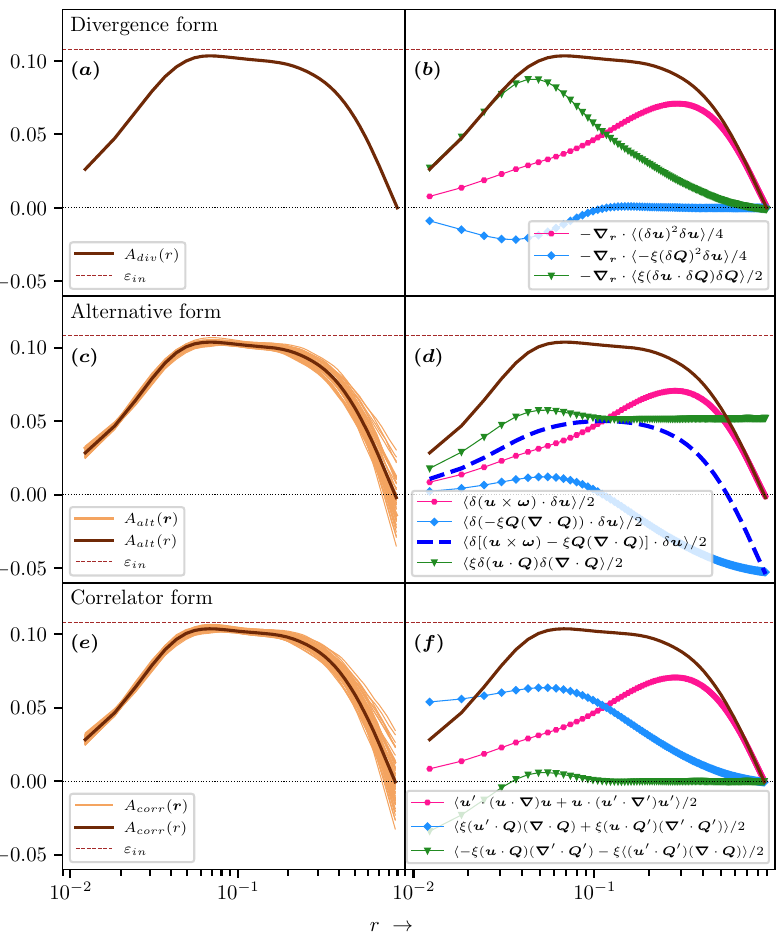}
    \caption{(a), (c), (e) (left panel) Plots of the total energy cascade rates obtained from the divergence form ($A_{div}(r)$), the alternative form ($A_{alt}(r)$), and the correlator form ($A_{corr}(r)$) of the exact relations, along with the direction-dependent cascade rates $A_{alt}(\bm{r})$ and $A_{corr}(\bm{r})$ for the alternative and correlator forms, respectively.
(b), (d), (f) (right panel): Term-by-term plots of the total energy cascade rate for the three exact laws for Run2.}
\label{fig:compare_all_div_alt_corr}
\end{figure}
\begin{figure}[ht!]
\includegraphics[width=0.85\linewidth]{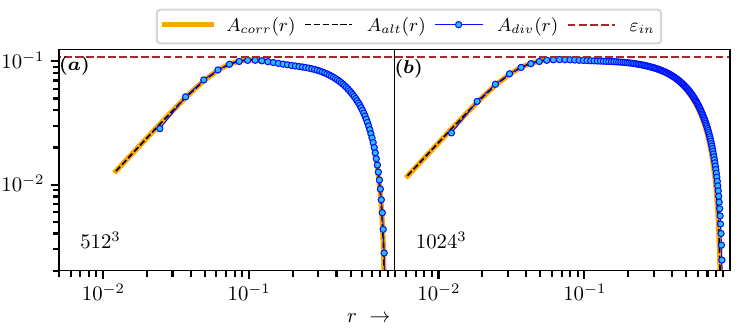}
    \caption{The overplot of the cascade rates obtained from the divergence form ($A_{div}(r)$), alternative form ($A_{alt}(r)$), and the correlator form ($A_{corr}(r)$) of the exact relations for Run1 ((a), left) and  Run2 ((b), right).}
    \label{fig:overplot_512_1024}
\end{figure}

\vspace{0.5cm}
\begin{centering}
    \subsection{Comparison of different exact relations}
\end{centering}
\vspace{-0.5cm}
  In the next step, we calculate the cascade rates $A_{alt}(\bm{r})$ and $A_{corr}(\bm{r})$ using the divergence-free forms of the exact laws, namely the alternative form (Eq.~\eqref{alt_form}) and the correlator form (Eq.~\eqref{corr_form}). 
  Unlike the divergence form, it is straightforward to calculate the direction dependent cascade rates $A_{alt}(\bm{r})$ and $A_{corr}(\bm{r})$ from Eqs.~\eqref{A_alt_r} and \eqref{A_corr_r}, respectively. However, to compare them with the isotropic $A_{div}(r)$, we also compute $A_{alt}(r) =  \sum A_{alt} (\bm{r})/73$ and $A_{corr}(r) =  \sum A_{corr} (\bm{r})/73$. In Fig.~\ref{fig:compare_all_div_alt_corr} (a), we plot $A_{div}(r)$ whereas in Figs.~\ref{fig:compare_all_div_alt_corr} (c) and (e), we plot $A_{alt}(\bm{r})$ and $A_{corr}(\bm{r})$ along with their angle-averaged values $A_{alt}(r)$ and $A_{corr}(r)$, respectively. Evidently, all the isotropic cascade rates are found to behave identically with $r$.
 Interestingly, across a range of intermediate scales, the direction dependent cascade rates (as well as their average) are fairly overlapping. This is due to the isotropic nature of the cascade which is a direct consequence of the absence of a mean composition gradient field. Again, unlike in Fig.~\ref{fig:A_divflux_diss_einj}, the flat part of neither of the isotropic cascade rates exactly touches the constant $\varepsilon_{in}$. This slight discrepancy is prominent as we plot the cascade rates in algebraic scales and can possibly be attributed to (i) a non-perfect stationary state where only an approximate balance between the average injection and dissipation rates is attained, and (ii) a non-vanishing coupling between the forcing and inertial length scales. Finally, for the sake of direct comparison, we overplot all the three cascade rates $A_{div}(r)$, $A_{alt}(r)$ and $A_{corr}(r)$ in Fig.~\ref{fig:overplot_512_1024} which clearly shows an excellent agreement between all the three forms for both the grid resolutions $512^3$ and 
 $1024^3$. Despite their different expressions, all the three forms were derived under the common assumption of homogeneity, and hence the stark similarity between them clearly indicates that the binary fluid turbulence remains highly homogeneous even in the presence of two-fluid domains and the chain-like structures in the $\bQ$ field.
 This is practically achieved owing to moderately high Weber numbers ($We \approx 12$ for $512^3$ and $We \approx 25$ for $1024^3$) which indicates the inertial forces to be dominating over the surface tension forces thereby causing significant breakdown of the individual fluid domains to form a homogeneous emulsion.  

\vspace{-.3cm}
\begin{centering}
    \subsection{Term-by-term investigation of different exact relations}
\end{centering}
\vspace{-.5cm}
 Proceeding further, we also study term-by-term contributions of the cascade rates obtained from the three forms of exact laws (see Fig.~\ref{fig:compare_all_div_alt_corr} (right panel)). In Fig.~\ref{fig:compare_all_div_alt_corr} (b), we observe that the inertial range $A_{div}(r)$ is mainly dominated by the term $\bnabla_{\bm{r}}\cdot\langle- (\delta \bu)^2\delta \bu \rangle$ (pink) for larger scales whereas the contributions, comprising of the $ \bQ$ field, become prominent at smaller scales.  In particular, at large scales, the injected kinetic energy efficiently cascades towards the smaller scales via the term $\bnabla_{\bm{r}}\cdot\langle- (\delta \bu)^2\delta \bu \rangle$ whereas the term $\bnabla_{\bm{r}}\cdot\langle -2 \xi (\delta \bu \cdot \delta \bQ)\delta \bQ\rangle$ (green) provides the average transfer rate of large scale kinetic energy to small scale active energy. Finally, the active energy stored at the interfacial scales undergoes an inverse cascade at a rate of $\bnabla_{\bm{r}}\cdot\langle \xi ( \delta \bQ)^2\delta \bu \rangle$ (blue) up to the scale $r \sim 0.1$. By phenomenological arguments, one can show this scale roughly corresponds to the Hinze-scale ($\sim (\sigma^3/\varepsilon^2)^{1/5}\sim 0.13$ in the present case) at which phase-arrested structures are formed due to an interplay between the flow and surface dynamics of a binary fluid.

  For the alternative form, it is easy to recognize $\langle\delta(\bu \times \bomega)\cdot\delta \bu\rangle$ (pink in Fig.~\ref{fig:compare_all_div_alt_corr} (d)) is exactly equal to the flux-rate for kinetic energy (pink in Fig.~\ref{fig:compare_all_div_alt_corr} (b)) and hence decreases as one approaches the smaller scales inside the inertial range. Together with the contribution $\langle\xi\delta(- Q(\bnabla\cdot\bQ))\cdot \delta \bu\rangle$ (blue curve Fig.~\ref{fig:compare_all_div_alt_corr} d), it provides the energy cascade rate due to the imbalance between $(\bu \times\bomega)$ and $\xi \bQ(\bnabla\cdot\bQ)$ which stays fairly constant (dashed curve in deep blue) across the inertial range. The term due to $\langle\xi\delta(\bu\cdot\bQ)\delta (\bnabla\cdot\bQ)\rangle$ (green) also remains constant for scales larger than the domain size whereas it increases slightly at scales comparable to the interface width. Interestingly, the sum of all the contributions finally gives a reasonably constant flux rate very similar to the that obtained from the divergence form. 

The kinetic part of the correlator form $\langle \bu'\cdot(\bu\cdot\bnabla)\bu + \bu\cdot(\bu'\cdot\bnabla')\bu'\rangle$ (pink curve in Fig.~\ref{fig:compare_all_div_alt_corr} (f)) is found to exactly match with that of $A_{alt}(r)$, whereas the other terms of $A_{corr}(r)$ do not show a term by term matching with $A_{alt}(r)$ (Fig.~\ref{fig:compare_all_div_alt_corr} (f)). However, a careful inspection shows that the blue and green curves of Fig.~\ref{fig:compare_all_div_alt_corr} (f) are simply obtained from the blue and green curves of Fig.~\ref{fig:compare_all_div_alt_corr} (d) by adding and subtracting the average one-point contribution $\langle 2 \xi(\bu \cdot\bQ)(\bnabla \cdot\bQ)\rangle$, respectively. For the pure kinetic part such a shift is not visible as $\langle\bu \cdot (\bu \times \bomega)\rangle$ vanishes identically. 
Again, the individual terms of $A_{div}(r)$ and $A_{alt}(r)$ are expressed in terms of two-point increments and hence vanish for $r \rightarrow 0$. This is no longer true for $A_{corr}(r)$ that consists of mixed two-point correlators that do not necessarily vanish at $r \rightarrow 0$. 

\vspace{0.5cm}
\begin{centering}
\section{Summary and conclusion} \label{Summary and conclusion}  
\end{centering}
\vspace{-0.5cm}
Despite a number of studies on high Reynolds number turbulence in binary fluids, dedicated studies on universal energy cascades were visibly lacking. It is only recently that we derived two exact relations in divergence and correlator forms for homogeneous BFT \citep{Pan_2022_Exact_BFT}. Built upon that foundation, the present work provides a comprehensive extension where we first derive a third alternative form of the exact relation. The alternative formulation explicitly characterizes the turbulent energy cascade as deviations from the Beltrami or force-free type of aligned states for HD and MHD systems \cite{Banerjee_2016_alternative}. In this case, the alternative form clearly implies that the presence of any of the following conditions $\bu||\bomega$,  
$\xi\bQ(\bnabla\cdot\bQ) = \bnabla A$, $\bu\perp\bQ$ $etc.,$ shall lead to partial suppression of turbulent cascade for binary fluid systems, where ${\bf A}$ can be any arbitrary vector field. 

The derivation is followed by an extensive numerical investigation of the energy cascade in fully developed BFT by means of those three forms of exact relations. Using direct numerical simulation of up to $1024^3$ grid points, we systematically demonstrate that the kinetic and active energies do not separately exhibit universal cascades across the inertial range. However, the total energy exhibits a Kolmogorov-like cascade with a constant flux rate equal to the average injection rate at largest scales. Furthermore, albeit with term by term disparities, the total energy cascade rates calculated from all the three forms are found to be exactly equal across length scales thus confirming the equivalence of the three forms of exact relations. Consequently, the turbulent heating rate for homogeneous BFT can be equally determined using any of these laws, depending on the analytical, computational, or experimental needs. The equivalence of these three forms also ensures the existence of a homogeneous turbulence in a phase arrested regime. \\
All the exact laws are derived without any prior assumption of isotropy, making them equally applicable to both isotropic and anisotropic flows, such as BFT with a strong mean composition gradient field. Such situations indeed arise in practical scenarios, including salinity variations in saltwater-freshwater systems, temperature gradients in oil-water or water-alcohol mixtures, and chemotaxis in chemical binary mixtures $etc.,$ \cite{Kim2022_Investigating_water_oil_interfaces, Stürmer_Chemotaxis_in_binary_mixture}. Unlike the divergence form, computing flux rates using the alternative and correlator forms is more straightforward in such cases, as it does not require any prior knowledge about the flow symmetry. However, the directional cascade rates will be no longer the same as that of the average cascade rate. A future study can be conducted to investigate the effect of the mean field on the energy cascade rate using all the three exact laws where it would be interesting to examine the mutual deviations among them as a function of the strength of mean field. In addition, in all the three forms, the kinetic and non-kinetic parts of the flux rates cross each other roughly at the domain size and a plausible phenomenological explanation is also provided for the same. However, a stand-alone study is essential to decipher the fundamental relation between the domain size and the crossing point of different flux rates. Finally, this study can also be extended to compressible and active binary fluids which are of surmount interest for chemical and biological sciences as well as industrial applications. 

\begin{centering} 
\vspace{-0.0cm}
\section{Acknowledgment}
\end{centering}
\vspace{-0.3cm}
The simulations are performed using the support and resources provided by PARAM Sanganak under the National Supercomputing Mission (NSM), Government of India at the Indian Institute of Technology, Kanpur. For data analysis, we again acknowledge NSM for providing computing
resources of ‘PARAM UTKARSH’ at CDAC Knowledge Park Bangalore, which is implemented
by C-DAC and supported by the Ministry of Electronics and Information Technology (MeitY)
and Department of Science and Technology (DST), Government of India.

We acknowledge Prasad Perlekar and  Samriddhi Sankar Ray for useful discussions. 
%

\end{document}